\documentclass[
 preprint, 
 amsmath,amssymb,
 aps, physrev,
floatfix,
twocolumn,
showkeys
]{revtex4-2}

\usepackage{graphicx}
\usepackage{dcolumn}
\usepackage{bm}
\usepackage{xcolor}
\usepackage{hyperref}
\usepackage{subfig}
\usepackage[mathlines]{lineno}

\usepackage[left=1.2cm,right=1.2cm,top=1.8cm,bottom=2cm]{geometry}

\begin{document}

\preprint{}

\title{\textbf{Probing new physics in the Boosted $HH \to b\bar{b}\gamma\gamma$ channel at the LHC} 
}%

\author{Mohamed Belfkir}
\email{m\_belfkir@uaeu.ac.ae}
\affiliation{
 Department of Physics, United Arab Emirates University, Al-Ain 15551, UAE
}%

\date{\today}

\begin{abstract}
This paper presents the first dedicated study of the boosted $HH \to b\bar{b}\gamma\gamma$ topology as a key probe of physics beyond the Standard Model (SM) in the high-energy double-Higgs boson regime. The analysis presented in this paper, focuses on two classes of new-physics scenarios: non-resonant deviations of the quartic gauge--Higgs interaction, parameterized by the coupling modifier $\kappa_{2V}$, and resonant enhancement arising from the decay of a heavy scalar state, modeled within a two-Higgs-doublet framework. We demonstrate that the boosted reconstruction category enhances sensitivity to beyond SM effects that populate the high-$m_{HH}$ tail, yielding improved constraints on 
$\kappa_{2V}$ and extending the discovery reach for heavy resonances.
\end{abstract}

\maketitle


\section{Introduction}
\label{sec:intro} 

Understanding the interactions that govern the Higgs sector remains a central challenge in particle physics. Although a decade of precision measurements has established the properties of the Higgs boson discovered at the LHC ~\cite{Aad_2012,Chatrchyan_2012,HIGGS1964132,PhysRevLett.13.321,PhysRevLett.13.508,Higgs_mass_ATLAS,Higgs_mass_CMS} to be compatible with the SM, the form of the Higgs potential and the structure of its couplings are still only loosely constrained. In particular, the trilinear Higgs self-coupling $\lambda_{3}$ which governs the shape of the Higgs potential and the interactions between the Higgs field and electroweak gauge bosons play a key role in electroweak symmetry breaking and in a wide range of extensions of the SM~\cite{PhysRevD.110.030001,PhysRevD.9.3357,Laine:471776,Li:2019jba,arXiv:2407.14716}.  

At the LHC proton-proton collisions, the Higgs boson pair production ($pp\to HH$) provides a unique laboratory to probe these interactions. While the dominant gluon–gluon fusion (ggF) process is mainly sensitive to the self-coupling modifier $\kappa_{\lambda}=\lambda_{3}/\lambda_{3}^{\text{SM}}$, the vector-boson fusion (VBF) production mode provides the unique direct test of the gauge–Higgs sector through the $VVH$ and $VVHH$ vertices. Many theoretical frameworks, including two-Higgs-doublet models, singlet extensions, and super-symmetric scenarios~\cite{Lee:1973iz,Branco:2011iw,Haber:2015pua,Chen:2013jvg,Muhlleitner:2016mzt,Chalons:2012qe,Robens:2019kga,Djouadi:2005gj,Gunion:1984yn,Gunion:1986nh,Degrassi:2002fi,Maniatis:2009re,Ellwanger:2009dp,King:2012is}, predict deviations in double Higgs boson production either through the new resonant decaying to two Higgs bosons or through deviations in the quartic gauge–Higgs interaction, commonly parameterized by the modifier $\kappa_{2V}$.  
The VBF channel is particularly sensitive to these deviations, especially in the high-energy regime where the two forward jets and the Higgs system exhibit large boost~\cite{Grazzini_2018,Baglio_2021,Dreyer_2018,Dreyer_2020,bbyy_legacy,ATLAS:2022jtk,ATLAS:2024ish,CMS_HH}. The search for new resonants in the double Higgs boson production and constraining $\kappa_{2V}$ has therefore emerged as a priority for Higgs physics program at the LHC. 

The $HH\to b\bar{b}\gamma\gamma$ channel provides one of the most experimentally powerful probes of Higgs pair production due to its relatively clean final state and high trigger efficiency compared to the fully hadronic channels. 
Despite its small branching ratio of roughly 0.26\% for $m_H=125\,$GeV~\cite{CERNYellowReportPageBR}, the final state combines the large $H\to b\bar{b}$ branching fraction with the exceptionally clean diphoton signature, where the narrow $m_{\gamma\gamma}$ peak enables strong suppression of the dominant non-resonant continuum $\gamma\gamma$ + jets background. Both ATLAS and CMS have exploited this final state extensively, setting stringent limits on both non-resonant and resonant $HH$ productions~\cite{bbyy_run2,bbyy_legacy,atlas_results,CMS:2024awa,Adhikary:2020fqf,Wu:2025jza}. All the $HH\to b\bar{b}\gamma\gamma$ analyses to date have operated in the fully \textit{resolved} reconstruction regime, where the two $b$-quarks emerging from $H\to b\bar{b}$ are reconstructed as separate small-radius jets.  
This configuration provides good sensitivity to the double Higgs boson production cross-section limit and to variations of $\kappa_{\lambda}$, and moreover has been the foundation for recent machine-learning–enhanced studies, including our earlier works employing graph neural networks~\cite{Belfkir:2025zlx} and hybrid quantum machine learning \cite{AitHaddou:2025hol}.  
However, the resolved topology probes moderate kinematic scales and thus has limited sensitivity to the boosted high-energy regime where the deviations in $\kappa_{2V}$ or new scalar resonance decaying to two Higgs bosons are expected. 

In the boosted regime, where the Higgs boson is produced with large transverse momentum ($p_T$), the decay products of $H\to b\bar{b}$ become highly collimated, yielding to a single large-radius jet with a distinctive two-prong substructure. Such \textit{boosted} Higgs boson configuration have become essential in analyses involving hadronic Higgs decays, specially for the $HH \to b\bar{b} b\bar{b}$~\cite{JetGNN,Guo:2020vvt,Hammad:2025aka}, offering improved reconstruction efficiencies and enabling access to extreme kinematic regions. Recently, the CMS Collaboration uses the boosted topology to exclude the $\kappa_{2V}=0$ scenario with a significance of 6.3~$\sigma$, demonstrating the strong discriminating power of the boosted phase space and further motivating dedicated studies in this regime~\cite{CMS:2022gjd}. The boosted category enables sensitivity to both modifications of $\kappa_{2V}$ and to heavy resonance decaying to Higgs boson pairs.

Despite the strong theoretical motivation, the $HH\to b\bar{b}\gamma\gamma$ channel has thus far been explored primarily in the resolved regime, and the potential of a dedicated boosted category has not yet been systematically assessed. In this paper, we present the first detailed study of the boosted $HH\to b\bar{b}\gamma\gamma$ topology for both non-resonant and resonant searches. Building on the simulation framework of Refs.~\cite{Belfkir:2025zlx, AitHaddou:2025hol}, we define two mutually exclusive analysis regimes: a resolved category, in which the $b$-jets from $H\to b\bar{b}$ are reconstructed separately as two small-radius jets, and a boosted category, where the Higgs boson decay is captured as a single large-radius jet. An XGBoost classifier is used in the non-resonant analysis to define signal-enriched score bins within each category. For the resonant search, the available statistics at high transverse momentum are limited, and the signal topology is already highly distinctive. Thus, a simple rectangular-cut selection is adopted instead of a machine-learning approach. The central goal of the study is not to optimize classification performance, but to demonstrate the importance of accessing the high-energy VBF-like phase space through boosted Higgs reconstruction in the $HH \to b\bar{b}\gamma\gamma$ channels. 

This paper is organized as follows. Section~\ref{sec:HH_intro} introduces the theoretical framework. The Monte Carlo event generation is described in Section~\ref{sec:mc}. Object reconstruction and event definition are detailed in Section~\ref{sec:obj_event}. The event selection strategy is presented in Section~\ref{sec:selection}. The results are presented and discussed in Section~\ref{sec:results}. Finally, the conclusions are given in Section~\ref{sec:con}.

\section{Theoretical Framework}
\label{sec:HH_intro}

\begin{figure}[ht]
    \centering
    \subfloat[a][ggF box]   {\includegraphics[width=0.35\linewidth]{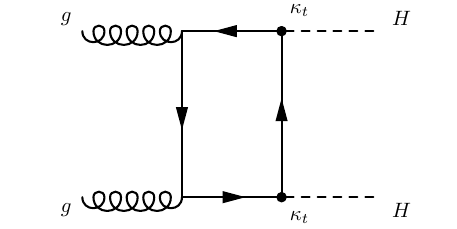}}\hspace{0.2cm}
    \subfloat[b][ggF triangle]{\includegraphics[width=0.3\linewidth]{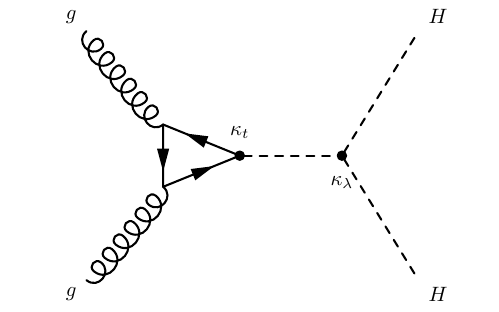}}\\
    \subfloat[c][VBF $\kappa_{\lambda}$]{\includegraphics[width=0.33\linewidth]{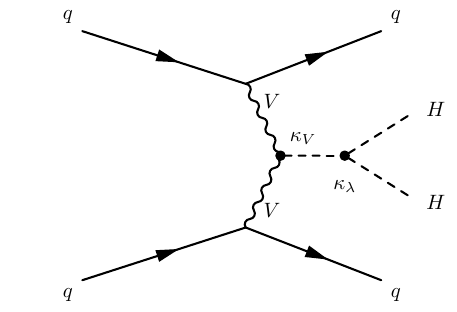}}\hspace{0.2cm}
    \subfloat[d][VBF $\kappa_{V}$]{\includegraphics[width=0.28\linewidth]{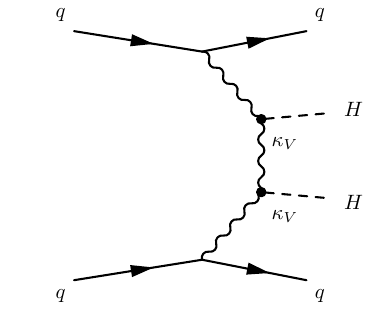}}\hspace{0.2cm}
    \subfloat[e][VBF $\kappa_{2V}$]{\includegraphics[width=0.28\linewidth]{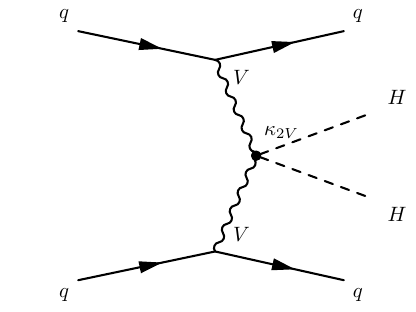}} \\
    \subfloat[f][Resonant]{\includegraphics[width=0.28\linewidth]{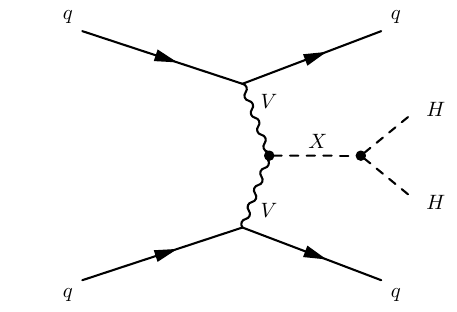}}
    \caption{Leading-order Feynman diagrams for (a,b) gluon–gluon fusion and (c–e) vector-boson-fusion Higgs boson pair production. (f) Illustrates the resonant production mode.}
    \label{fig:feyn}
\end{figure}

Higgs boson pair production in proton--proton collisions proceeds primarily through the ggF production mode, with a smaller but theoretically important contribution from VBF production mode. The leading-order diagrams contributing to ggF and VBF production are shown in Figure~\ref{fig:feyn}.

In the non-resonant case, the ggF process receive contribution from two leading amplitudes: the \emph{box} diagram (Figure \ref{fig:feyn}~a) and the \emph{triangle} diagram (Figure \ref{fig:feyn}~b).  
The latter depends explicitly on the trilinear Higgs self-coupling and interferes destructively with the box contribution, making the total cross section strongly dependent on the modifier $\kappa_{\lambda}$.  
The VBF production mode probes a different sector of the Higgs potential. In addition to $\kappa_{\lambda}$, its amplitude receives contributions from both the $VVH$ and $VVHH$ vertices, making the process sensitive to modifications of the Higgs–vector boson couplings $(\kappa_V)$ and, critically, to the quartic gauge–Higgs interaction parameterized by $\kappa_{2V}$.  
Modifications in the $\kappa_{2V}$ alter the high-energy behavior of the VBF amplitude and affect both the inclusive rate and the kinematic distributions of the Higgs boson pair.

In the resonant case, many theories predict the existence of heavy scalar states that can decay 
into a pair of Higgs bosons, producing a resonant enhancement in the $HH$ invariant-mass spectrum
\cite{Branco:2011iw, Barman:2020ulr}.  At leading order (Figure~\ref{fig:feyn}~f), the resonant production mechanism corresponds to a heavy state $X$ being produced through VBF, followed by the decay $X \to HH$, yielding kinematics characterized by two highly boosted Higgs bosons whenever $m_{X} \gg 2~m_{H}$. In this analysis, we consider the presence of such a resonant as a benchmark to illustrate the sensitivity of the boosted $HH\to b\bar{b}\gamma\gamma$ topology to heavy beyond SM (BSM) particles. For simulation purposes, we adopt a simplified scenario inspired by the type-II two-Higgs-doublet model (2HDM) in which the SM-like Higgs $H$ is accompanied by a heavier neutral scalar $X$ that decays dominantly 
into $HH$. No model-dependent constraints, parameter scans, or couplings are used in the interpretation; the 2HDM framework serves only as a generator-level tool to produce representative kinematics for different scalar resonant masses $m_{X}$.  As the resonant mass increases, the two Higgs bosons recoil back-to-back with large transverse 
momentum, and their hadronic decays become increasingly collimated.  This places the bulk of the signal in precisely the region where the resolved selection loses 
efficiency and the boosted reconstruction becomes essential.

\begin{figure}[ht]
    \centering
    \subfloat[VBF (SM)][VBF (SM)]{\includegraphics[width=0.5\linewidth]{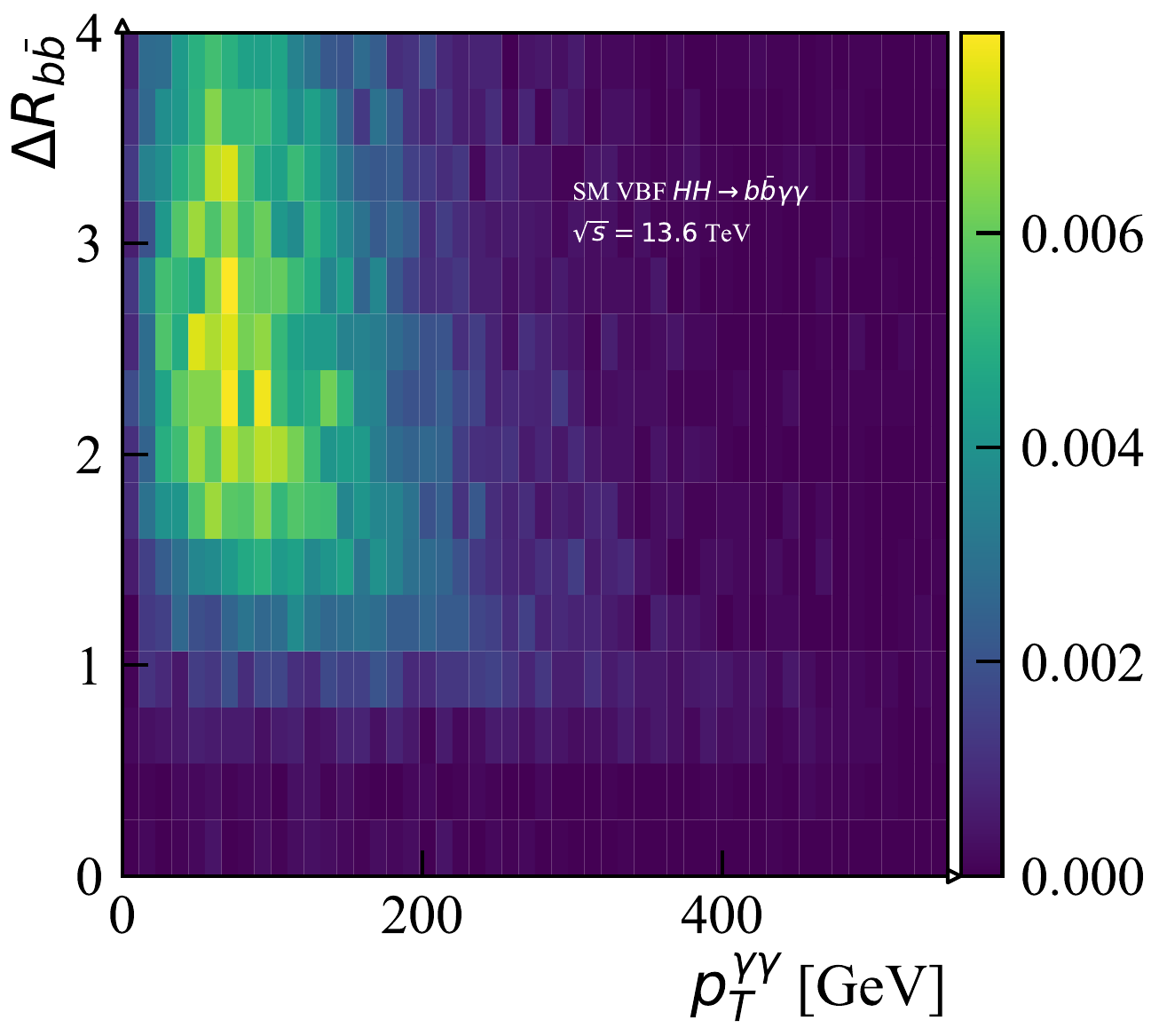}}
    \subfloat[VBF ($\kappa_{2V}$ = 0)][VBF ($\kappa_{2}V$ = 0)]{\includegraphics[width=0.5\linewidth]{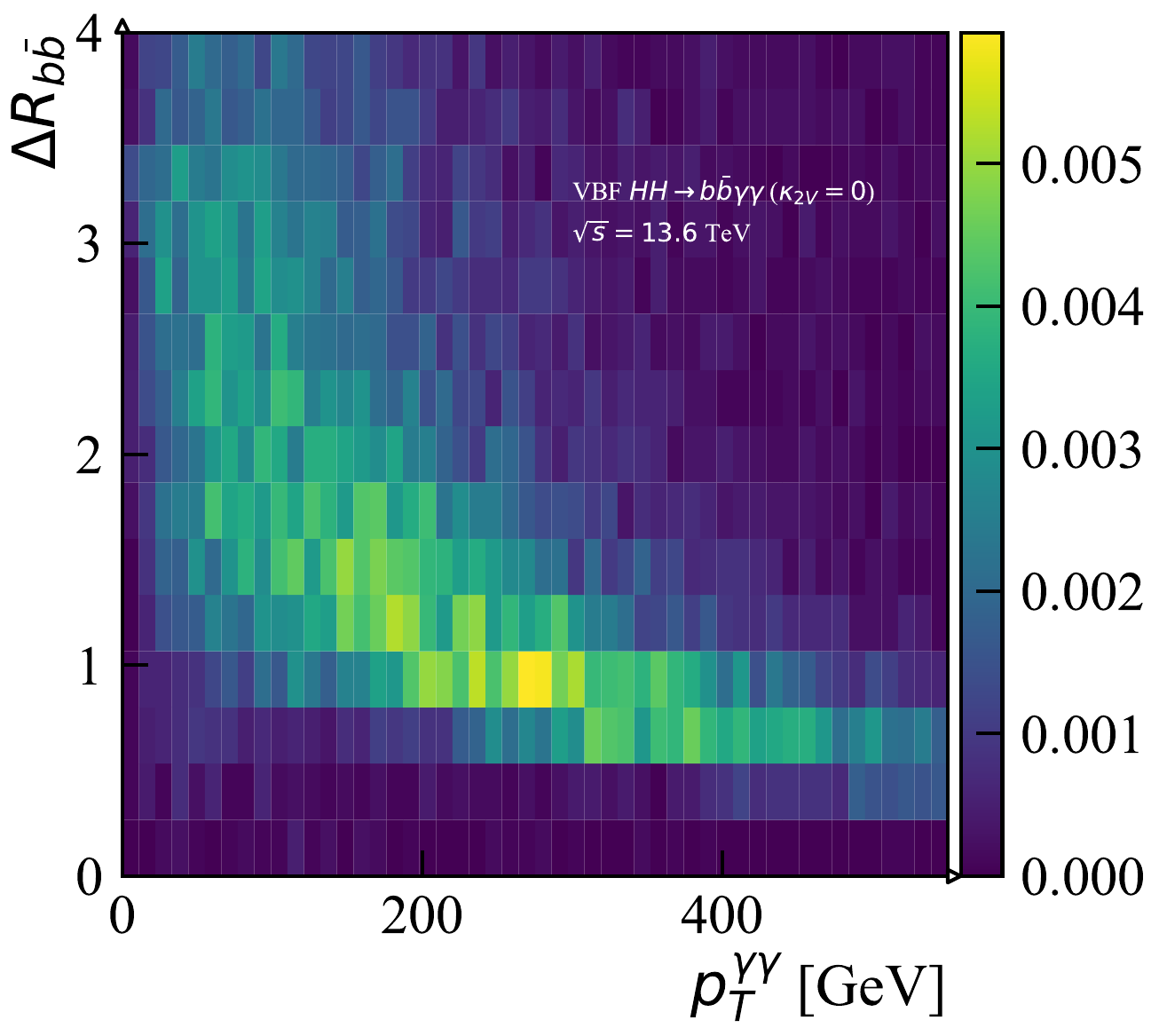}} \\

    \subfloat[$m_{X}$ = 2 TeV][$m_{X}$ = 2 TeV]{\includegraphics[width=0.5\linewidth]{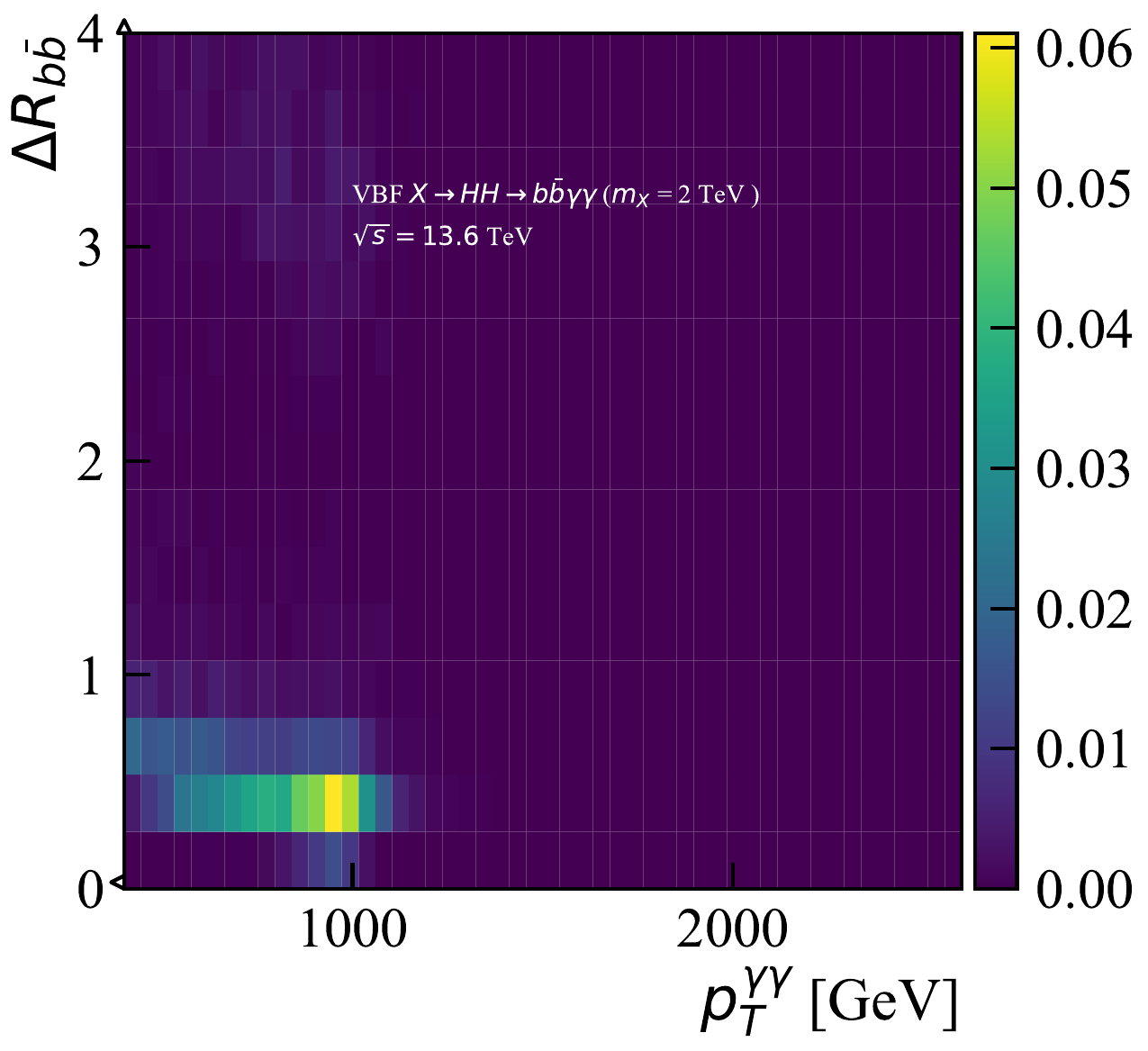}}
    \subfloat[$m_{X}$ = 5 TeV][$m_{X}$ = 5 TeV]{\includegraphics[width=0.5\linewidth]{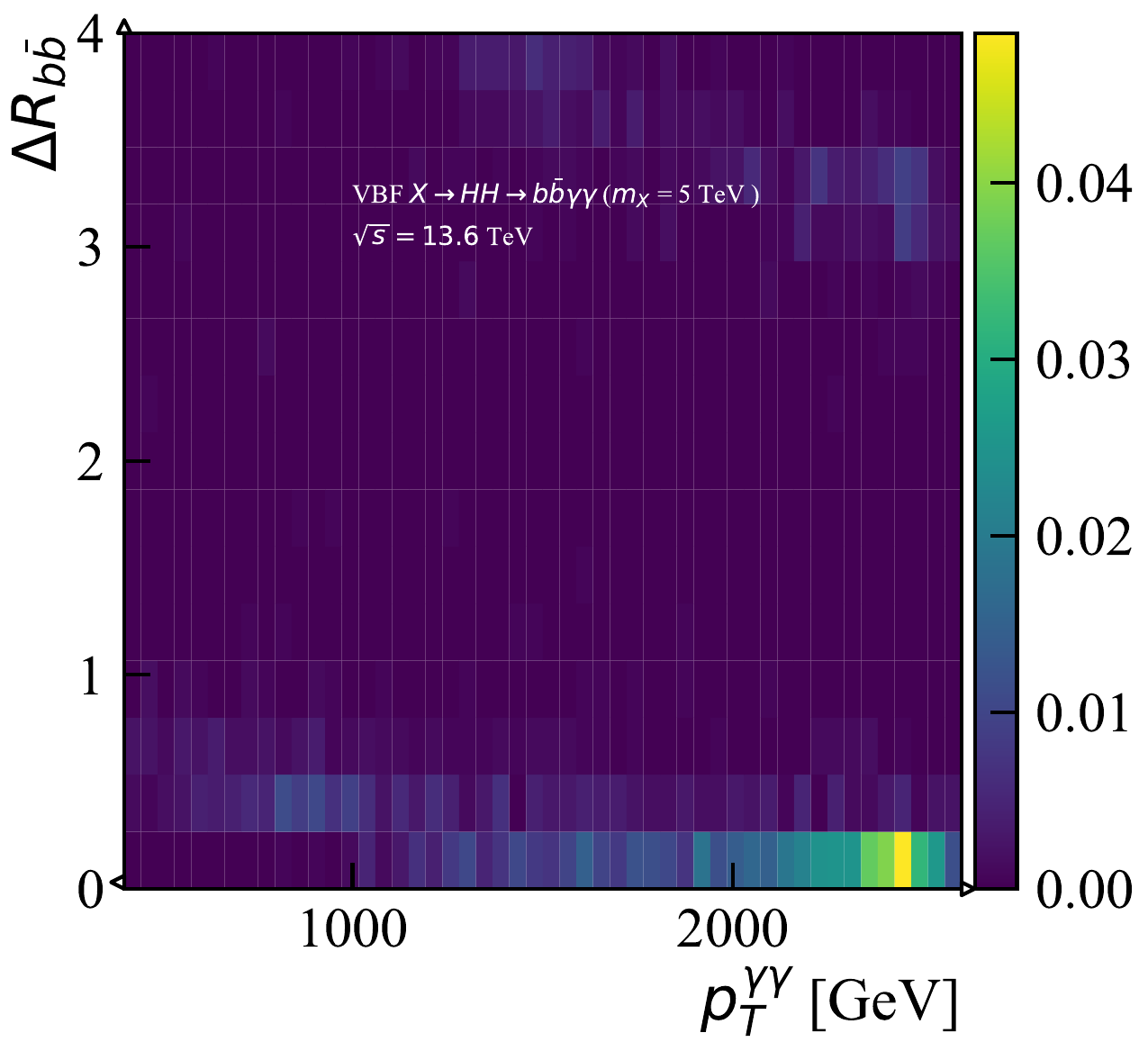}}
    
    \caption{Two-dimensional truth-level distribution of  
        $\Delta R_{b\bar b}$ as a function of $p_{T}^{\gamma\gamma}$ for SM (a), $\kappa_{2V}=0$ (b), resonant $m_X$ = 2~TeV (c) and $m_X$ = 5~TeV (d) at $\sqrt{s}=13.6$~TeV.}
    \label{fig:2DdR}
\end{figure}

Figure \ref{fig:2DdR} shows the two-dimensional distribution of the angular separation 
between the two $b$ quarks, $\Delta R_{b\bar b}$, as a function of the diphoton transverse momentum $p_{T}^{\gamma\gamma}$ at the truth-level. At moderate Higgs transverse momenta such as for SM (Figure \ref{fig:2DdR}~a), the two $b$ quarks from $H\to b\bar{b}$ are well separated, typically with $\Delta R_{b\bar{b}} \gtrsim 0.4$, and are reconstructed as two distinct small-radius (small-$R$) jets. This defines the resolved regime, which has been the focus of most existing experimental and phenomenological studies~\cite{bbyy_run2,bbyy_legacy,atlas_results,CMS:2024awa,Adhikary:2020fqf,Wu:2025jza,Belfkir:2025zlx,AitHaddou:2025hol}. However, we focus on the \emph{semi-boosted} regime (hereafter referred to simply as the boosted regime), in which only the $H\to b\bar{b}$ decay products become collimated while the di-photon system remains well resolved. In this configuration, the characteristic angular separation of the $b$ quarks decreases approximately as  
\begin{equation}
\Delta R_{b\bar{b}} \simeq \frac{2m_{H}}{p_{T}^{H}}.
\end{equation}
Once $\Delta R_{b\bar{b}} \lesssim R_{\text{jet}}$, both $b$ quarks are captured inside a single large-radius (large-$R$) jet, as shown in Figure~\ref{fig:2DdR} for BSM scenarios. The resulting jet exhibits a distinctive two-prong substructure, a feature that has been extensively exploited in a variety of hadronic Higgs analyses~\cite{JetGNN,Guo:2020vvt}.

\section{Monte Carlo Event Generation}
\label{sec:mc}

The Monte Carlo (MC) samples used in this analysis correspond to simulated proton--proton collisions at a center-of-mass energy of $\sqrt{s}=13.6~\text{TeV}$. The simulation setup is designed to model the kinematics of Higgs boson pair production in the $H(\to b\bar{b})H(\to\gamma\gamma)$ final state and to provide sufficient statistical precision to evaluate the performance of the boosted and resolved reconstruction strategies introduced in this work. Both signal and background samples are normalized to the most accurate cross-section predictions available in the literature. An integrated luminosity of $308~\mathrm{fb}^{-1}$ is assumed, representative of the combined Run~2 and partial Run~3 dataset recorded by the LHC experiments. 

For the non-resonant signal, the ggF process is generated using the \texttt{Powheg-Box~v2} framework \cite{Alioli_2010}, including finite top-quark mass effects and full NLO QCD corrections.  
The generated events are subsequently normalized to the NNLO cross-section prediction with top-mass dependence following the Refs.~\cite{Grazzini_2018,Baglio_2021}.  
The VBF signal sample is produced at the leading order using \texttt{MadGraph5\_aMC@NLO~v3.3.0} \cite{Alwall_2014}, and the normalization is set using the N3LO QCD + NLO electroweak prediction 
from Refs.~\cite{Dreyer_2018,Dreyer_2020}. 
All signal samples are generated with the Higgs boson mass fixed to $m_H = 125~\text{GeV}$ and SM couplings ($\kappa_{\lambda}$ = 1, $\kappa_{2V}$ = 1). Additional VBF samples are generated with varied $\kappa_{2V}$ parameter from its SM value including $\kappa_{2V} = 0, 0.5, 1.5, -1 ~\text{and}~-2.5$. The VBF resonant BSM samples are generated at the leading order using \texttt{MadGraph5\_aMC@NLO~v3.3.0} using the 2HDM-Type-II model for eight mass point $m_{X}$ = 1, 1.5, 2, 2.5, 3, 3.5, 4 and 5 TeV. The generated samples are normalized to a cross-section equal to 1 fb.

Backgrounds contributing to the $b\bar b\gamma\gamma$ final state are classified into resonant and non-resonant components. Resonant backgrounds originate from single Higgs boson production followed by the decay $H\to\gamma\gamma$. The dominant production mechanisms--ggF, VBF, $ZH$, and $t\bar{t}H$--are generated with \texttt{Powheg-Box~v2}, using identical settings to those employed for the ggF signal simulation. The cross-sections are taken from the most recent LHC Higgs Working Group recommendations~\cite{CERNYellowReportPageBR}. The dominant non-resonant background arises from direct QCD production of continuum $\gamma\gamma$ events accompanied by jets. This background is produced with \texttt{MadGraph5\_aMC@NLO}, including matrix elements with up to two additional partons in the final state to ensure an accurate modeling of the jet multiplicity and the kinematic tails relevant for this analysis.  Table~\ref{tab:mc_table} summarizes the generated processes, along with 
the corresponding numbers of simulated events and nominal cross-sections.

\begin{table}[htb]
    \centering
    \begin{tabular}{lcc}
    \hline\hline
     Process  & Events & Cross-section [pb] \\
     \hline
     ggF $HH$ (SM)  & 50k & 0.3413 \cite{Dreyer_2020}\\
     VBF $HH$ (SM)  & 100k & 0.001874 \cite{Dreyer_2020}\\
     VBF $HH$ ($\kappa_{2V} = 0$)  & 100k & 0.02885 \cite{Dreyer_2020} \\
    \hline 
    VBF $X \to HH$& 100k & 1 fb \\
     \hline
     ggF $H$  & 500k & 52.17 \cite{CERNYellowReportPageBR}\\
     VBF $H$  & 550k & 4.075 \cite{CERNYellowReportPageBR}\\
     $qq\to ZH$  & 50k & $1.834\times10^{-3}$ \cite{CERNYellowReportPageBR}\\
     $gg\to ZH$  & 100k & $3.087\times10^{-4}$ \cite{CERNYellowReportPageBR}\\
     $t\bar t H$  & 100k & 5.688$\times10^{-1}$ \cite{CERNYellowReportPageBR}\\
     \hline
     $\gamma\gamma$ + jets & 2.5M & 48.1 \\
     \hline\hline
    \end{tabular}
    \caption{Summary of the generated signal and background samples used in this analysis, together with their nominal cross-sections.}
    \label{tab:mc_table}
\end{table}

All parton-level events are processed through \texttt{Pythia~8.186}~\cite{pythia8} for parton showering, hadronization, and underlying-event modeling. The detector response is emulated using the \texttt{Delphes} fast-simulation framework~\cite{delphes}, configured with an ATLAS detector card tuned for Run~3 performance.

\section{Object reconstruction and Event selection}
\label{sec:obj_event}

The final state targeted in this analysis contains two isolated photons and the hadronic decay of a Higgs boson into a $b\bar b$ pair. Depending on the transverse momentum of the hadronically decaying Higgs, the two $b$ quarks may be reconstructed either as two distinct small-$R$ jets (resolved topology) or as a single large-$R$ jet  (boosted topology).  All reconstructed objects are obtained from the \texttt{Delphes} fast simulation of the ATLAS detector and follow Run~3 object definitions.

\subsection{Object reconstruction}
\label{sec:obj}

Photon candidates are built from clusters in the electromagnetic calorimeter and are required to satisfy a transverse momentum threshold of $p_{T} > 20~\text{GeV}$ and a pseudorapidity acceptance of $|\eta|<2.37$.  The region $1.37<|\eta|<1.52$ is excluded to avoid reduced calorimeter response.  A dedicated emulation of the ATLAS Run~3 photon  identification efficiency is applied using the efficiency maps provided in Ref.~\cite{photonID}. When two photon seeds are nearly coincident in $(\eta,\phi)$ space ($\Delta R < 0.01$), only the higher-$p_T$ photon is retained, ensuring the isolation of photon candidates.

Small-$R$ jets used in the resolved analysis are formed using the anti-$k_t$ algorithm with radius parameter $R$ = 0.4, as implemented in 
\texttt{FastJet}~\cite{Cacciari_2008,Cacciari:2011ma}. Reconstructed jets are required to satisfy $p_{T}>25~\text{GeV}$ and $|y|<4.5$. A parameterized $b$-tagging response corresponding to 85\% 
efficiency~\cite{btagGN2} is applied to identify jets originating from $b$-hadrons. This working point corresponds to mis-tag rates of 0.17 for $c$-jets and 0.01 for light-flavor jets. Only $b$-tagged small-$R$ jets within the inner-detector coverage ($|\eta|<2.5$) are considered.

To capture merged $H\to b\bar b$ decays in the boosted regime, a second jet collection is constructed using the anti-$k_t$ algorithm with $R=1.0$. To reflect modern ATLAS grooming and substructure practices, several procedures are applied at the Delphes level to improve the stability of the large-$R$ jet mass and suppress contamination from soft radiation and pile-up \cite{Krohn:2009th}. Each jet is first trimmed by reclustering its constituents into subjets of size $R_{\mathrm{trim}}=0.2$ and removing subjets carrying less than 5\% of the original jet $p_T$. Pruning is then applied using $z_{\mathrm{cut}}=0.1$ and $R_{\mathrm{cut}}=0.5$ to eliminate soft, large-angle contributions. In addition, the jet undergoes soft-drop grooming with $\beta_{\mathrm{SD}}=0$, a symmetry threshold of $z_{\mathrm{cut}}=0.1$, and characteristic radius $R_{0}=0.8$, providing a groomed mass that is less sensitive to soft QCD effects \cite{Larkoski:2014wba,Moult:2016cvt}. Large-$R$ jets used as boosted Higgs candidates are required to satisfy a transverse momentum in the range $250~\text{GeV} < p_{T} < 3~\text{TeV}$ and a rapidity acceptance of $|y| < 2$, ensuring that the jet is fully contained within the inner detector tracker. After grooming, the jet mass is required to lie withing $50 < m_{J} < 600~\text{GeV}$, a broad window chosen to retain signal-like two-prong configurations while suppressing soft QCD backgrounds. Identification of jets originating from a merged $b\bar{b}$ system is performed using a parametrized double-$b$ tagging efficiency of 75\%, with a corresponding mis-tag probability of approximately 6\% for light-flavor and gluon-initiated jets. 

Electrons and muons are reconstructed from the particle-flow track collection. They are used exclusively for vetoing backgrounds with leptonic activity. Electrons must satisfy $p_{T}>10~\text{GeV}$ and $|\eta|<2.47$, excluding the 
ATLAS barrel--endcap transition region. Muon candidates are required to have $p_{T}>10~\text{GeV}$ and $|\eta|<2.7$. 

To ensure a unique assignment of reconstructed objects and to prevent double counting of calorimeter deposits or tracks, a fixed overlap–removal sequence is applied. Jets that lie within $\Delta R<0.4$ of any reconstructed photon are discarded, and jets overlapping with electrons within $\Delta R<0.2$ are also removed. After the jet cleaning step, electrons found within $\Delta R<0.4$ of any remaining jet are removed, as are muons satisfying the same condition. Photon candidates overlapping with electrons or muons within $\Delta R<0.4$ are subsequently discarded, guaranteeing that photons used in the analysis are isolated from charged leptons. This procedure yields mutually exclusive collections of jets, photons, and leptons, providing a clean object definition for the resolved and boosted reconstruction strategies used in this study.

\subsection{Common Event selection}
\label{sec:common}

All events considered in this analysis must satisfy a diphoton trigger designed to select final states 
containing two energetic photons. The trigger thresholds correspond to transverse energy ($E_T$) requirements of approximately 35~GeV for the leading photon and 25~GeV for the subleading one, and the associated efficiencies are applied to the simulated events using the measured trigger efficiency obtained in ATLAS Run~3 data at $\sqrt{s}=13.6~\text{TeV}$~\cite{trig}.  
Only events that pass this trigger emulation and the object selections described in 
Sec.~\ref{sec:obj} are retained.

To form a Higgs boson candidate decaying into two photons, the two highest-$p_{T}$ photons in the event 
are selected and required to satisfy an invariant mass window,
\[
105~\text{GeV} < m_{\gamma\gamma} < 160~\text{GeV},
\]
which defines the signal-enriched diphoton region. In addition, scaled $p_T$ requirements are imposed to ensure a uniform selection efficiency across the diphoton mass range. Specifically, the leading photon must satisfy $p_{T}^{\gamma_1} > 0.35\,m_{\gamma\gamma}$, and the subleading photon must meet
$p_{T}^{\gamma_2} > 0.25\,m_{\gamma\gamma}$. These normalized thresholds suppress events with soft photons and preserve the shape of the $m_{\gamma\gamma}$ spectrum near the Higgs boson mass, preventing distortions that would otherwise complicate the background modeling~\cite{ATLAS:2014euz}. The photon pair passing these criteria defines the reconstructed $H\to\gamma\gamma$ candidate.

Events containing reconstructed electrons or muons are vetoed to suppress contributions from 
top-associated and electroweak processes.  
Furthermore, to reduce contamination from hadronic $t\bar{t}H$ backgrounds, events with more than six 
central jets are rejected. This set of requirements defines the common selection applied to both the resolved and boosted categories. 

\subsection{Resolved Category Selection}

In the resolved topology, the hadronic Higgs boson candidate is reconstructed from two well-separated 
$b$-tagged small-$R$ jets. Events entering this category must contain at least two jets identified as originating from $b$-quarks, and the pair with the highest transverse momenta is taken to represent the 
$H\to b\bar b$ decay.  No additional constraints are imposed on the angular separation of the two jets beyond the requirements already included in the object definitions; the resolved configuration is naturally 
selected by the absence of collimation between the two $b$ quarks. To characterize the VBF event topology, additional jets not used in the Higgs reconstruction are examined. If such jets are present, the two highest-$p_T$ remaining jets are designated as VBF candidates and may be used to probe the forward-jet kinematics typical of VBF production mode.  
This identification is auxiliary and not required for an event to be accepted: events without a pair of 
suitable forward jets are kept in the sample, ensuring that the resolved selection remains inclusive and highly sensitive to the SM cross-section limit.

\subsection{Boosted Category Selection}

The boosted category targets events in which the hadronic Higgs decay is reconstructed as a single large-$R$ jet. The highest-$p_{T}$ large-$R$ jet meeting the criteria described in Section~\ref{sec:obj} is selected as the Higgs $H\to b\bar{b}$ candidate. The boosted category is required to have at least two VBF jets. To identify a VBF-like topology, the remaining small-$R$ jets in the event are examined.  Only jets that are well separated from the Higgs candidate, $\Delta R(j, J) > 1.4$, are considered. Among all possible jet pairs fulfilling this condition, the pair with the largest pseudorapidity separation ($|\Delta\eta|$) is chosen as the VBF-jet system. 

\subsection{Orthogonality}

To ensure orthogonality between the two analysis categories, in events where a suitable large-$R$ jet is identified, the hadronic Higgs decay is represented by this single boosted jet, and the event is treated accordingly. If no such jet is found, but the event contains two distinct $b$-tagged small-$R$ jets consistent with a resolved $H\to b\bar b$ configuration, the reconstruction follows the resolved approach instead. This procedure assigns each event to one topology only, with boosted candidates taking priority whenever they are present, thereby preventing any ambiguity between the two reconstruction modes. The overlap fraction between the two categories is found to be 8.7\% for the $\kappa_{2V}$ = 0 and 2.2\% for the SM sample.



\begin{figure}[ht]
    \centering
    \subfloat[non-resonant analysis][non-resonant analysis]{\includegraphics[width=0.5\linewidth]{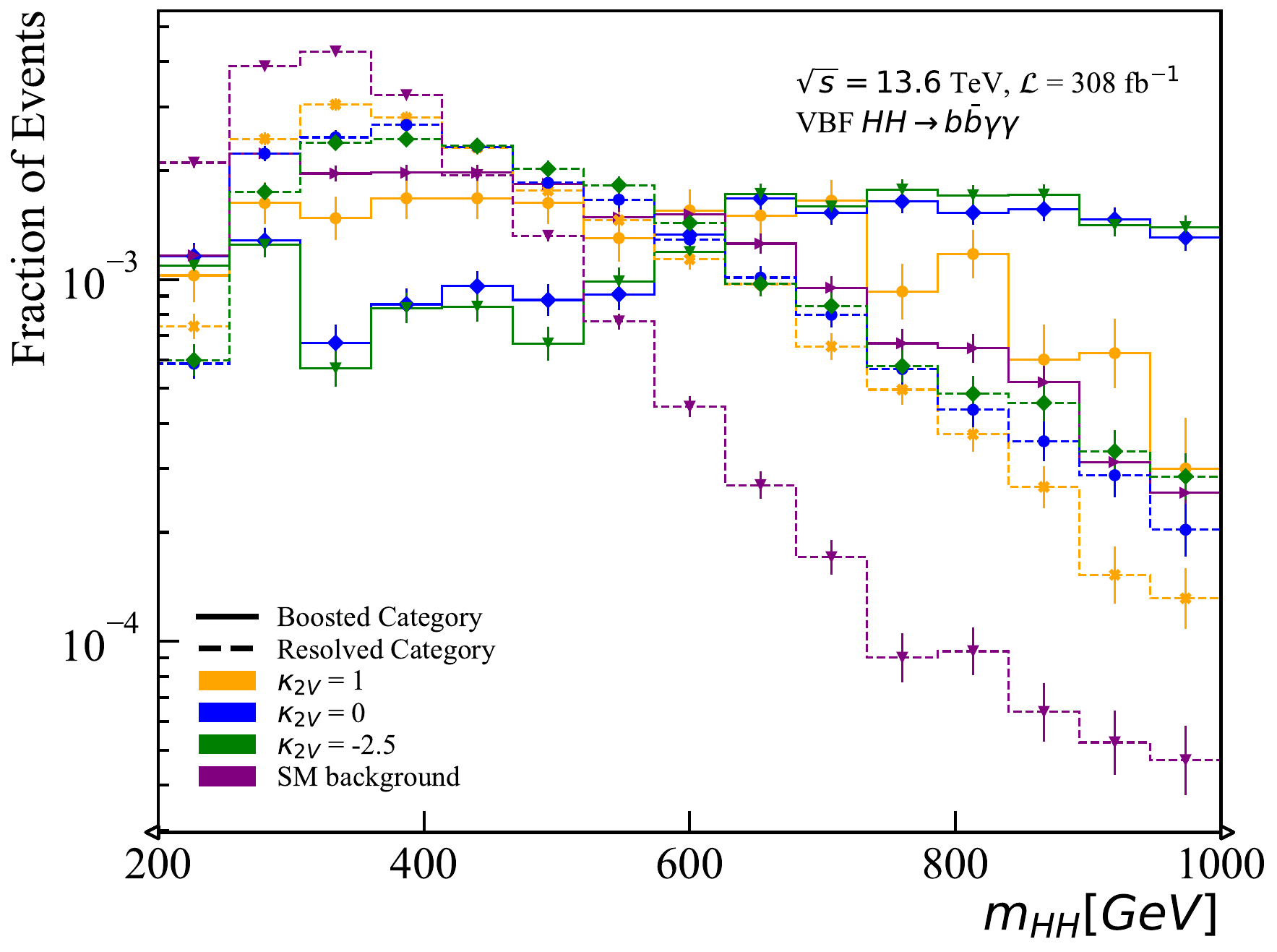}}
    \subfloat[resonant analysis][resonant analysis]{\includegraphics[width=0.5\linewidth]{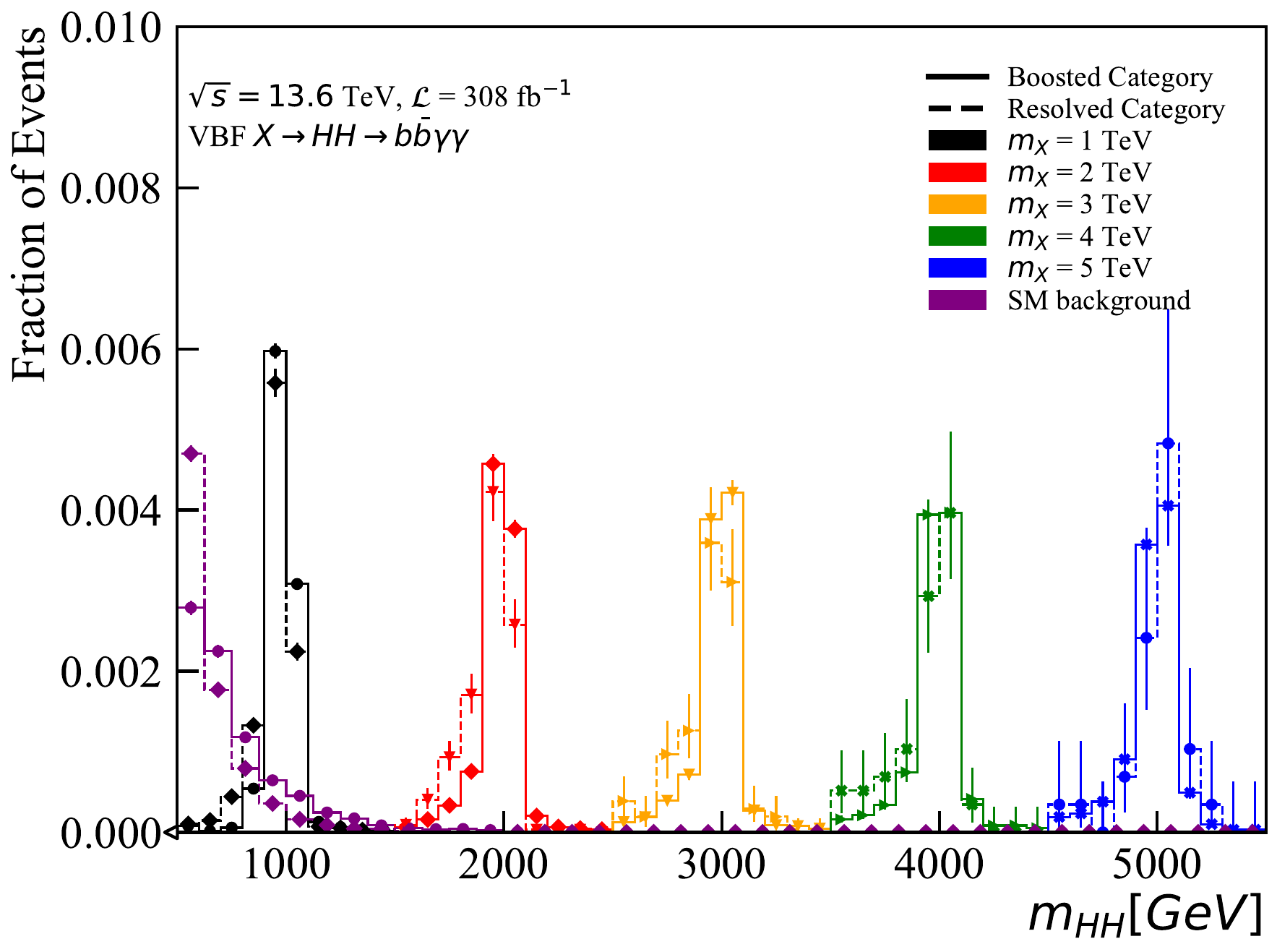}}
    \caption{The reconstructed di-Higgs invariant mass $m_{HH}$ in both non-resonant (a) and resonant (b) analysis. The boosted reconstruction category is shown with solid histograms, while the resolved category is shown with dashed histograms. }
    \label{fig:mhh}
\end{figure}

Figure \ref{fig:mhh} shows the reconstructed di-Higgs invariant mass distributions for (a) the non-resonant analysis and (b) the resonant analysis after applying the full event selection described above. The distributions are shown separately for the boosted reconstruction category (solid histograms) and the resolved reconstruction 
category (dashed histograms). In the non-resonant case, the boosted category demonstrates an enhanced sensitivity to deviations in $\kappa_{2V}$ in the high-$m_{HH}$ tail, where contributions from BSM effects are expected to be largest. A similar behavior is observed in the resonant analysis, where the boosted reconstruction dominates the sensitivity at large resonant masses $m_{X}$.

\section{Event categorization}
\label{sec:selection}

This section describes the strategy used to categorize events and construct a lightweight classifier aimed at improving the separation between signal and background for the non-resonant study. The emphasis of the analysis is not on developing an advanced ML model, but rather on demonstrating how a dedicated boosted reconstruction of the $HH \to b\bar b \gamma\gamma$ final state can strengthen the sensitivity to BSM physics—an aspect that has not been systematically explored in previous studies. Due to the lack of background events in the high-$p_T$ $m_{HH}$ spectrum, the resonant analysis will relay only on the rectangular cuts described before and no ML based categorization will be considered.  

Before training the XGBoost classifier \cite{xgboost}, a uniform pre-processing procedure is applied to all selected events to ensure consistent inputs and to reduce detector-related asymmetries. In particular, events that satisfy either of the Higgs reconstruction paths (resolved or boosted) are transformed by rotating the final-state momenta in the transverse plane so that the leading photon is placed at a fixed azimuthal reference direction. This operation leverages the approximate cylindrical symmetry of the ATLAS detector and removes arbitrary variations in $\phi$, enabling the classifier to concentrate on physically relevant geometric correlations rather than on random orientation differences~\cite{Chatham_Strong_2020}.  
All remaining objects in the event are rotated coherently to preserve their internal angular relationships. For the purposes of training and validating the classifier, the full event sample is partitioned into two non-overlapping subsets.  
A randomly selected fraction of 75\% is used to optimize the model parameters, while the remaining 25\% is held out and serves exclusively for performance validation. All physics results presented in this work rely solely on the latter subset, ensuring that the quoted performance is derived from events not used during training and is therefore free from training-induced bias. Each simulated event carries a weight proportional to its generator-level cross section, allowing the effective event composition to reproduce the expected yields in data.  
Because the background samples greatly exceed the signal in raw statistics, an additional per-class normalization factor is applied to counteract the imbalance. 

Separate XGBoost classifiers are trained for the resolved and boosted event categories. In the resolved case, the signal sample consists of the SM ggF and VBF Higgs boson pair production.  
For the boosted classifier, the signal definition is extended to include events generated with 
$\kappa_{2V}=0$ and $\kappa_{2V}=0.5$, providing additional statistics in the kinematic region where 
merged $H\to b\bar{b}$ decays are most frequent. Both classifiers rely on a common set of high-level kinematic variables describing the diphoton system, 
including $p_T$, $\eta$, and $\phi$ angles of the two leading photons, as 
well as the four-vector properties of the reconstructed $H_{\gamma\gamma}$ candidate 
($p_{T}^{\gamma\gamma}$, $\eta^{\gamma\gamma}$, $\phi^{\gamma\gamma}$, and $m_{\gamma\gamma}$).  
Kinematic quantities associated with the VBF-tagged jets are also included: the $p_T$, 
$\eta$, and $\phi$ angle of each jet, together with the di-jet angular separation 
$\Delta\eta_{jj}$ and invariant mass $m_{jj}$. The two categories differ only in the representation of the hadronic Higgs candidate. In the resolved regime, the classifier receives as input the four-vector components of the two 
$b$-tagged small-$R$ jets, the corresponding reconstructed $H_{b\bar b}$ system 
($p_{T}^{b\bar b}$, $\eta^{b\bar b}$, $\phi^{b\bar b}$, and $m_{b\bar b}$) and the reconstructed HH system ($p_{T}^{b\bar b\gamma\gamma}$, $\eta^{b\bar b\gamma\gamma}$, $\phi^{b\bar b\gamma\gamma}$, and $m_{b\bar b\gamma\gamma}$).  
In the boosted regime, the hadronic Higgs is instead identified with the leading large-$R$ jet, and the 
classifier uses the jet kinematics ($p_{T}^{J}$, $\eta^{J}$, $\phi^{J}$, and $m^{J}$) as the 
$H\to b\bar{b}$ representation and the $\gamma\gamma$ + large$R$ jet kinematic ($p_{T}^{J\gamma\gamma}$, $\eta^{J\gamma\gamma}$, $\phi^{J\gamma\gamma}$, and $m_{J\gamma\gamma}$) as the representation of the HH system in the boosted regime. The XGBoost algorithm includes a set of hyper-parameters \cite{xgboost-parameters} that significantly influence model performance and require careful tuning. In this analysis, both XGBoost classifier rely on the random grid search approach for the hyperparameter optimization.

The output of each XGBoost model is used to define two signal-enriched bins above the 0.5 threshold, with the bin boundaries optimized to maximize the expected sensitivity using the Asimov approximation to the profile likelihood ratio~\cite{Cowan:2010js} defined as: 
\begin{equation*}
    Z = \sqrt{2 \times ((s + b) \times \ln(1 + s / b) - s)},
\end{equation*}
where $s$ is the total signal yield and $b$ corresponds to the total background yield. The defined bins are used in the binned likelihood fit to compute the results.
  
\section{Results and Discussion}
\label{sec:results}

The purpose of this study is to highlight the sensitivity gain obtained by introducing a dedicated boosted $HH\to b\bar{b}\gamma\gamma$ category, particularly in the high mass region where many BSM scenarios predict enhancements. Two classes of new-physics effects are examined: (i) non-resonant deviations from the quartic gauge--Higgs interaction, parameterized through variations of $\kappa_{2V}$, and (ii) heavy resonances decaying into Higgs boson pairs, as predicted in extended scalar sectors. Both mechanisms increase the population of events with highly boosted Higgs bosons, making the boosted reconstruction highly sensitive over the resolved.

\subsection{Non-resonant analysis}

A binned likelihood fit is performed in each analysis category using the optimized classifier output bins.  
The statistical inference relies on the \texttt{pyhf} framework~\cite{Heinrich:2021gyp} to derive the 95\% confidence-level (CL) upper limit on the signal strength,  
\[
\mu_{HH} = \frac{\sigma(pp \to HH)}{\sigma_{\text{SM}}(pp \to HH)}.
\]
A 10\% background normalization uncertainty is included as a log-normal nuisance parameter. The observed data constructed by assuming the SM expectation (background plus SM ggF+VBF HH yields).  Because the resolved and boosted categories are strictly orthogonal, the combined likelihood of the two categories factorizes as  
\[
\mathcal{L}_{\text{combined}} =  
\mathcal{L}_{\text{resolved}} \times \mathcal{L}_{\text{boosted}}.
\]

\begin{figure}[htb]
    \centering
    \includegraphics[width=0.75\linewidth]{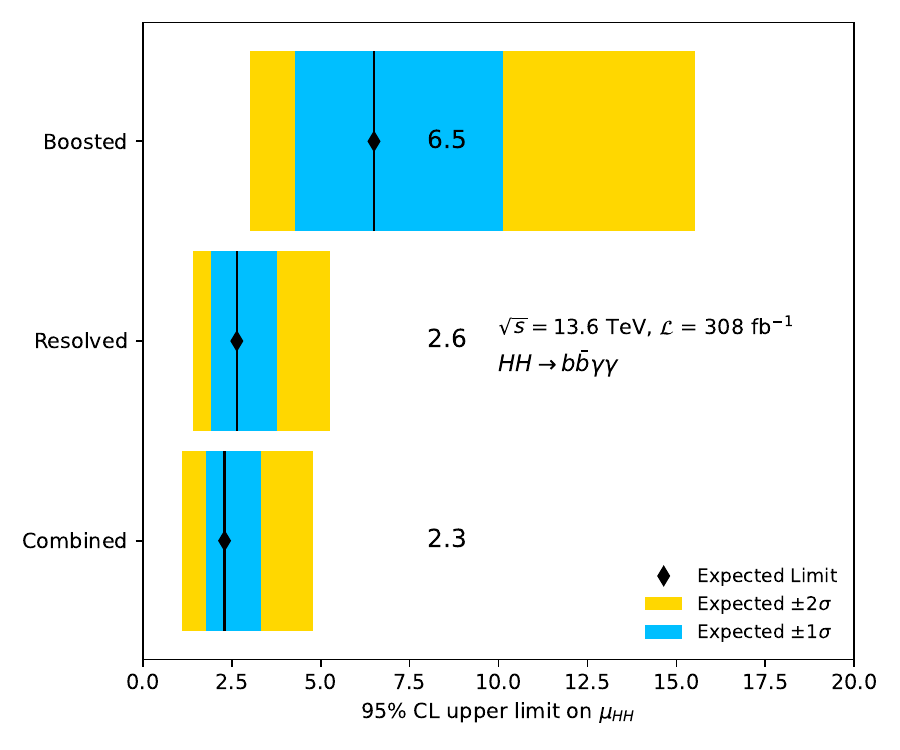}
    \caption{Expected 95\% CL upper limit on the signal strength $\mu_{HH}$ for the resolved, boosted, and combined categories.}
    \label{fig:limit}
\end{figure}

Figure~\ref{fig:limit} shows the extracted limits for each category.  The resolved channel provides the strongest individual constraint, excluding a signal strength of  
$\mu_{HH} = 2.6$ at 95\% CL, consistent with its high efficiency near threshold where the SM contribution peaks.  
The boosted channel yields a weaker limit of $\mu_{HH} = 6.5$, reflecting the much smaller SM contribution in the high-mass region.  When combined, the analysis achieves a limit of 2.3 times the SM predictions, dominated by the resolved region but benefiting from the complementary boosted acceptance.

\begin{figure}[htp]
    \centering
    \includegraphics[width=0.75\linewidth]{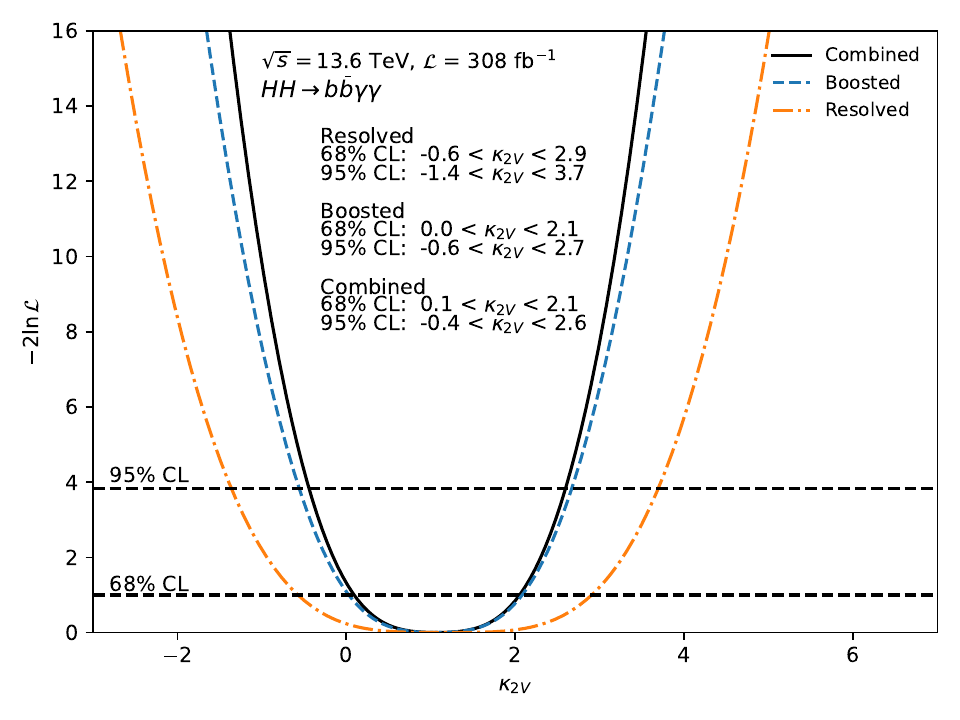}
    \caption{Profile-likelihood scan of the $\kappa_{2V}$ modifier for the resolved, boosted, and combined analysis categories.}
    \label{fig:k2v_scan}
\end{figure}

While the inclusive SM sensitivity is dominated by the resolved selection, the primary goal of this work is to probe deviations at high $m_{HH}$ induced by variations of the $\kappa_{2V}$ coupling in the non-resonant scenario. A one-dimensional profile-likelihood scan of $\kappa_{2V}$ is performed using the same statistical model as above. Here, only the impact of $\kappa_{2V}$ on the overall event rate is considered, without modeling shape modification.  Figure~\ref{fig:k2v_scan} shows the resulting likelihood profiles for the boosted, resolved, and combined categories.  
The boosted selection exhibits a pronounced response to non-SM values of $\kappa_{2V}$, significantly outperforming the resolved category in the large-$|\kappa_{2V}-1|$ region.  
This behavior reflects the strong correlation between $\kappa_{2V}$ variations and the high-$p_T$ phase space where the boosted topology dominates. The combined result inherits the strengths of both categories, providing superior constraints across the full parameter range.

\subsection{Resonant analysis}

Since the resonant analysis does not rely on the ML classifier, the likelihood is binned on the $m_{\gamma\gamma}$ in the range [120, 130] GeV with a bin width of 2 GeV. Similarly to the non-resonant analysis, the 95\% CL upper limit is derived on the signal strength, 
\begin{equation*}
   \mu = \frac{\sigma_{VBF}(X \to HH)}{1 ~\text{fb}}  
\end{equation*}
with a 10\% background normalization uncertainty and the observed data is constructed in the same way. The 95\% CL upper limit is evaluated for the eight generated mass points and a quadratic interpolation is used to interpolate between the mass point in the range 1 to 5 TeV. 

\begin{figure}[htp]
    \centering
    \includegraphics[width=0.7\linewidth]{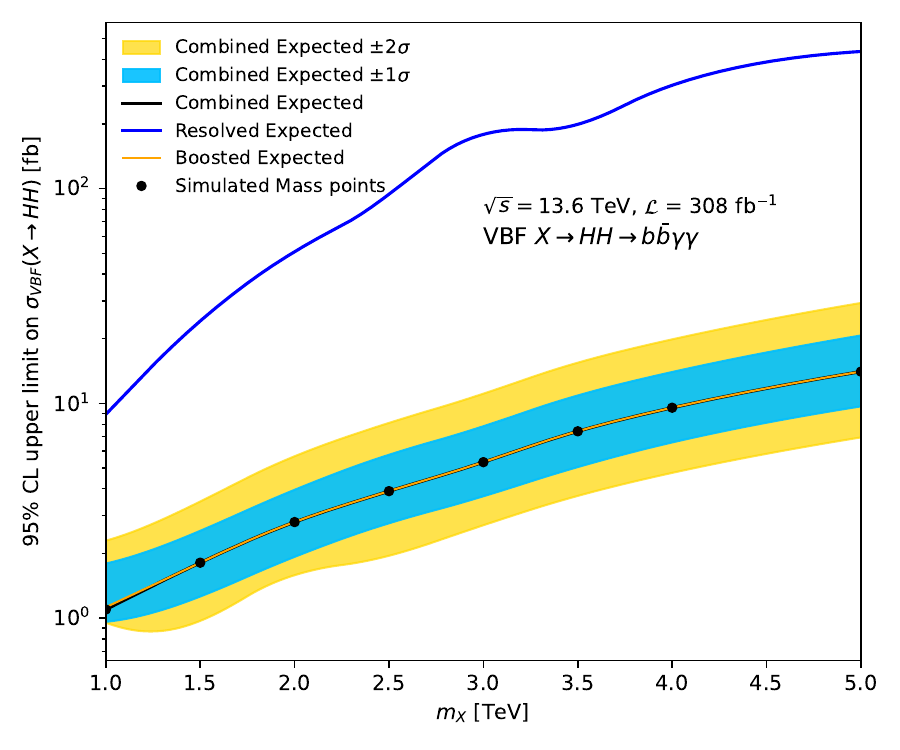}
    \caption{Expected 95\% CL upper limits on the VBF production cross section 
$\sigma(X \to HH)$ as a function of the resonant mass $m_{X}$ for the 
resolved (blue), boosted (orange), and combined (black) analysis categories. The combined expectation 
band includes the $\pm 1\sigma$ and $\pm 2\sigma$ variations, while the points 
represent the simulated mass hypotheses.}
    \label{fig:res_limit}
\end{figure}

Figure \ref{fig:res_limit} shows the 95\% CL upper limit on the VBF cross-section for the resonant process as a function of the heavy scalar mass $m_{X}$ for boosted, resolved and the combined cases. In the whole considered mass range, the boosted category provides significantly stronger limits than the resolved selection. This effect originates from the fact that resonant production manifests as enhancement to the high–$p_{T}$, high–$m_{HH}$ phase space where the  Higgs boson decay products are highly collimated, whereas the resolved regime suffers from vanishing acceptance once the two $b$-quarks merge. Consequently, for all mass points the limits set by the resolved category are well above the ones obtained with the boosted selection. The boosted category is able to set a limit on the cross-section about a factor one to two better depending on the mass point. This reflects the key role played by the boosted analysis in BSM searches. The combined limit is highly dominated with the boosted analysis setting limit ranging from 1 fb for 1 TeV and 100 fb at 5 TeV.

\section{Conclusion}
\label{sec:con}

This work presents the first dedicated analysis of the boosted 
$HH \to b\bar{b}\gamma\gamma$ topology, with a specific focus on understanding the relevance of these processes with regards to BSM theories that can extend the high energy phase space region. Both non-resonant BSM signals associated with modifications to the quartic couplings involving the gauge and Higgs boson, $\kappa_{2V}$ modifier, and resonantly produced Higgs pairs via heavy scalar decay ($X$), are considered. Both scenarios preferentially generate regions with large Higgs transverse momentum and high di-Higgs mass, with the hadronic Higgs system being reconstructed as large R jets instead of two separate $b$-jets.

To exploit this kinematic regime, two mutually exclusive and orthogonal reconstruction categories are defined: a resolved category conserving the sensitivity to SM Higgs boson pair production, and a boosted category—introduced here for the first time in the $b\bar{b}\gamma\gamma$ final state— targeting BSM-induced high-energy enhancements. The results demonstrate that the boosted category enhances the sensitivity of the golden di-Higgs channel to new physics effects that modify the high-$m_{HH}$ tail of the spectrum. In the non-resonant case, the boosted selection exhibits a strong dependence on $\kappa_{2V}$ away from the SM value, providing constraints that are not accessible with the resolved category alone. For heavy scalar resonances, the boosted reconstruction retains high acceptance across the mass range considered, while the resolved selection rapidly loses efficiency once the two $b$-quarks become collimated.

Consequently, it becomes clear that the resolved and boosted analysis combined greatly enhances the search region beyond what would have been achieved via a resolved-only analysis, establishing the $b\bar{b}\gamma\gamma$ channel as a competitive and complementary probe for BSM di-Higgs searches. This channel uniquely benefits from an efficient trigger strategy, well-understood backgrounds, and a clean experimental signature.

Overall, this study establishes the boosted $HH \to b\bar{b}\gamma\gamma$ topology as an essential tool for the BSM searches for the future Run-3 and HL-LHC physics program.

\begin{acknowledgments}
This work is supported by the United Arab Emirates University (UAEU) Start-Up Grant No 12S157. The authors gratefully thank the AI and Robotics Lab of United Arab Emirates University for offering computing facilities including HPC and DGX1 for MC simulation and ML training.
\end{acknowledgments}
\section*{Data and Code Availability}
The datasets and code used in this analysis can be provided by the corresponding author upon reasonable request.\\

\bibliography{apssamp}%
\end{document}